\documentclass[aps,twocolumn,showpacs,floatfix,superscriptaddress,prb]{revtex4-1}
\usepackage[utf8x]{inputenc}
\usepackage{graphicx}

\usepackage{amsmath}
\usepackage{fixmath}
\usepackage{amssymb}
\usepackage{hyperref}

\newcommand{\vn}[1]{\boldsymbol #1}

\newcommand{\xb}{{\boldsymbol x}}
\newcommand{\yb}{{\boldsymbol y}}

\begin{document}

\title{Phase Transition in $3d$ Heisenberg Spin Glasses with
  Strong Random Anisotropies, through a Multi-GPU Parallelization}
\author{M. Baity-Jesi} \affiliation{Departamento de F\'\i{}sica Te\'orica I,
  Universidad Complutense, 28040 Madrid, Spain.}  \affiliation{Instituto de
  Biocomputaci\'on and F\'{\i}sica de Sistemas Complejos (BIFI), 50009
  Zaragoza, Spain.}  \affiliation{Dipartimento di Fisica, Universit\`a La
  Sapienza, 00185 Roma, Italy.}

\author{L.A. Fern\'andez}
\affiliation{Departamento de F\'\i{}sica Te\'orica I, 
  Universidad Complutense, 28040 Madrid, Spain.}
\affiliation{Instituto de Biocomputaci\'on and
  F\'{\i}sica de Sistemas Complejos (BIFI), 50009 Zaragoza, Spain.}

\author{V. Mart\'in-Mayor}
\affiliation{Departamento de F\'\i{}sica Te\'orica I, 
  Universidad Complutense, 28040 Madrid, Spain.}
\affiliation{Instituto de Biocomputaci\'on and
  F\'{\i}sica de Sistemas Complejos (BIFI), 50009 Zaragoza, Spain.}

\author{J.M. Sanz}
\affiliation{Departamento de F\'\i{}sica Te\'orica I, 
  Universidad Complutense, 28040 Madrid, Spain.}

\date{\today}

\begin{abstract}
\noindent
We characterize the phase diagram of anisotropic Heisenberg spin glasses,
finding both the spin and the chiral glass transition.  We remark the presence
of strong finite-size effects in the chiral sector.  On the spin glass sector,
we find that the Universality class is that of Ising spin glasses.  Our data
is compatible with a unique phase transition for the chiral and spin glass
sector. We focus on keeping finite-size effects under control, and we stress
that they are important to understand experiments.  Thanks to large GPU
clusters we have been able to thermalize cubic lattices with up to $64^3$
spins, over a vast range of temperatures (hence, of relaxation times).
\end{abstract}


\maketitle

\section{Introduction}
\label{sec:Introduction}
Spin Glasses (SG) are disordered magnetic
alloys.\cite{Virasoro,Young97,Mydosh} Their microscopic modelization includes
several interactions, such as the RKKY interaction that is invariant over
rotations,\cite{Ruderman54,Kasuya56,Yosida57} and the Dzyaloshinsky-Moriya
(DM) interaction that breaks the rotational
symmetry.\cite{Dzyaloshinsky58,Moriya60} Therefore, in theoretical physics
SGs are often studied with simplified models that take in account
only a few essential characteristics (in particular, quenched disorder and
symmetries).\cite{EA75}

The DM interaction, through a spin-orbit coupling with a third spin, causes
the interactions between spins in any SG to have a certain degree of random
anisotropy. This implies that real SGs are never fully isotropic (this
theoretical limit is named Heisenberg SG).  In fact, materials are classified
according to the degree of anisotropy in their interactions,\cite{Petit02}
which turns out to be relevant in their non-equilibrium magnetic
response.\cite{Bert04} On one end of the materials' spectrum we find the
extremely anisotropic Fe$_{0.5}$Mn$_{0.5}$TiO$_3$, which is maybe the best
realization of the ideal limit of an Ising SG (Ising SGs correspond to the
idealization of uniaxial spins). On the other end, we have very isotropic
alloys such as AgMn or CuMn (whose modelization is notoriously
difficult,\cite{Peil09} due to the presence of short range spin-density wave
ordering\cite{Cable82,Cable84,Lamelas95}).

Despite the variety of interactions, already in the early '90s there was
general experimental agreement on that SGs undergo a phase transition at
sufficiently low temperature. \cite{Bouchiat86,Levy88,Gunnarsson91}

On the other hand, theoretical work was less advanced, even though one works
with extremely simple models. For the Ising SG there were arguments supporting
the existence of a phase transition,\cite{Franz94} that were later confirmed
numerically.\cite{Palassini99,Ballesteros00} In the Heisenberg case, instead,
all the attempts carried out during the '80s and '90s failed in finding a
phase transition at a finite temperature
$T_\mathrm{SG}>0$.\cite{McMillan85,Olive86,Morris86,Matsubara91} In fact,
Matsubara et al. argued in 1991 that once a small anisotropic term is added to
the Heisenberg Hamiltonian the phase transition becomes
visible.\cite{Matsubara91} This was in agreement with a later domain-wall
computation.\cite{Gingras93} The accepted picture at the time was that the
lower critical dimension (i.e. the spatial dimension below which there is no
phase transition) lies somewhere between 3D and 4D.\cite{Coluzzi95}

However, the story was slightly more complicated. Villain and coworkers made a
provocative suggestion hypothesizing that, although maybe there was no spin
glass transition, a different order parameter called chirality (or vorticity)
could be critical.\cite{Mauger90} Chirality is a scalar observable that
describes vorticity and alignment between neighboring spins (see below the
precise definition in Sect.~\ref{sec:obs}). This idea was elaborated by
Kawamura in his 1992 \emph{spin-chirality decoupling scenario:\/} in the ideal
case of a purely isotropic system the spin and chiral glass order parameters
would be decoupled, but the introduction of any small anisotropy would couple
them.\cite{Kawamura92}

Kawamura's scenario was apparently consistent with all the observations until
2003, when Lee and Young employed more efficient simulation algorithms and
finite-size scaling techniques to show that the spin glass channel is critical
also in the fully isotropic model (i.e. the Heisenberg limit).\cite{Lee03}
Both order parameters seemed to become positive at the same temperature.
Further simulations confirmed the existence of a spin glass phase transition,
although uncertainty remains on whether the transition is unique
\cite{Campos06,Fernandez09} or chiralities order at a slightly higher
temperature $T_\mathrm{CG}$.\cite{Viet09}

A parallel issue is measuring the chiral order parameter in
experiments. Kawamura proposed in 2003 that the extraordinary Hall resistivity
is a simple function of the linear and non-linear chiral-glass (CG)
susceptibilities.\cite{KawamuraExperiment03} Experiments based on this
proposal observed the chiral transition and measured, for instance, the
critical exponent $\delta$.\cite{Taniguchi07} Interestingly enough, the value
of $\delta$ turned out to be compatible between spin and
chiral glass sector. Nonetheless, it was impossible to identify a Universality
class despite the critical exponents of these systems had been extensively
measured (at least in the SG sector):\cite{Bouchiat86,Levy88, Petit02} the
impression was that they change in a continuous way from the Heisenberg to the
Ising limit, \cite{Campbell09} as we increased the anisotropy.

However, analogy with ferromagnetic materials suggests a different
interpretation. Anisotropy would be a relevant parameter in the sense of the
Renormalization Group.\cite{Amit05} There should be a new dominant fixed
point, and symmetry considerations lead to think it should belong to the
Ising-Edwards-Anderson Universality class. Yet, when we add a relevant
parameter to the Hamiltonian, there should be some \emph{cross-over}
effects. In other words, one expects that while the correlation length $\xi$
is small, the critical exponents are closer to the Heisenberg-Edwards-Anderson
Universality class, and that only for large enough $\xi$ the Universality
class reveals its nature.

Notwithstanding, it is very hard, both numerically and experimentally, to
prepare a SG with a large correlation length, since one should wait very long
times (it has been argued that the waiting time $t_\mathrm{w}$ required to
reach a certain coherence length is proportional to almost its seventh power,
see e.g. Refs. \onlinecite{Belletti08} and \onlinecite{Joh99}).  To our
knowledge, for this reason, the largest measured correlation lengths are of
the order of only one hundred lattice spacings. \cite{Joh99,Bert04} That is a
rather small distance to reveal the true Universality class, so it is
plausible that experiments will find critical exponents between the two
Universality classes.

To further complicate things, in experiments one has to take in account at
least two relevant crossovers. The first is the competition, that we just
pointed out, between the isotropic and the anisotropic fixed points. It is the
one we treat in this paper.  The second crossover, that we will not address, is
about short versus long range interactions.  In fact, the Hamiltonian we treat
is short range, but the DM interaction has been shown to be quasi-long-range,
in the sense that the interactions are long range, but only until a cut-off
distance of the order of some tens of atomic spacings. \cite{Bray82}

Aiming to untangle these questions, one of the authors undertook a numerical
study of Heisenberg SGs with very weak random anisotropies,
\cite{MartinPerez11} but the scenario remained even more foggy, since it was
observed that:
\begin{itemize}
\item The chiral glass critical temperature $T_{\mathrm{CG}}$ was
  significantly higher than $T_\mathrm{SG}$, in disagreement with experiments
  and expectations.
\item Apparently, the chiral susceptibility was \emph{not} divergent at
  $T_\mathrm{CG}$. This is surprising and, apparently, in contrast with
  experiments. \cite{Taniguchi07} Technically, this lacking divergence
  appeared as a very large anomalous dimension $\eta_\mathrm{CG}\sim
  2$. \cite{scalingRelation}
\item Introducing very weak anisotropies changed dramatically
  $T_\mathrm{SG}$. For example, the $T_\mathrm{SG}$ found by comparing
  systems of size $L=6,12$ was about twice its equivalent on the fully
  isotropic model. This is surprising, since one expects that the critical
  temperature would change very little from the isotropic case when $D$ is as
  small as in Ref. \onlinecite{MartinPerez11}.
\end{itemize}

In this paper we will focus on the uniqueness of the phase transition and on
the Universality class, proposing that there is a unique transition, belonging
to the Ising-Edwards-Anderson (IEA) Universality class.\cite{EA75} We will
also give an interpretation to the results of Ref. \onlinecite{MartinPerez11},
showing that the apparent inconsistencies are due to scaling corrections, that
we will try to characterize, since we believe them to be fundamental both in
the interpretation of numerical simulations and of experiments.

To do all this, we will study numerically the Heisenberg spin glass model with
strong random anisotropies, in order to suppress both Finite-Size Effects, and
traces of the cross-over from the isotropic limit.

We simulated on the largest lattices to present (up to $L=64$), over a wide
temperature range.\cite{Timerange} This has been possible thanks to an intense
use of graphic accelerators (GPUs) for the computations.  We made use of the
\emph{Tianhe-1A} GPU cluster in Tianjin, China,\cite{nscc} and of the
\emph{Minotauro} GPU cluster in Barcelona.\cite{bsc}

The rest of the paper is organized as follows. In Section \ref{sec:model} we
give an explicit definition of the model, and we introduce the observables we
extracted from simulations and analysis. Section \ref{sec:SimulationDetails}
contains details on how we practically conducted the simulations, although
much information is relegated to Appendix \ref{sec:numericaldetails}, where we
also discuss the use of GPUs for spin glasses. On Section \ref{sec:FSS} we
recall some Finite-Size scaling concepts we used in our analysis, to find the
critical temperatures and exponents (some technical details are given in
Section \ref{sec:methodology}). Finally, in Section \ref{sec:results} we refer
the results obtained in this work, and give our conclusions in Section
\ref{sec:conclusion}.

\section{Model and Observables}
\label{sec:model}

\subsection{The Model and its symmetries}\label{sec:model-symmetries} 

We study the model introduced by Matsubara et al.,\cite{Matsubara91} which is
particularly convenient because of its simplicity. We consider $N=L^3$
3-dimensional unitary vectors $\vec s_\xb = (s_\xb^1, s_\xb^2, s_\xb^3)$ on a
cubic lattice of linear size $L$, with periodic boundary conditions. The
Hamiltonian is 
\begin{equation}\label{eq:HamiltonianDefinition}
  H = -\sum_{<\xb,\yb>} ( J_{\xb\yb} \vec s_\xb \cdot\vec s_\yb +
  \sum_{\alpha\beta} s^\alpha_\xb D^{\alpha\beta}_{\xb\yb} s^\beta_\yb),
\end{equation}
where $<\cdot>$ means the sum goes only over nearest neighbors, and
the indices $\alpha,\beta$ indicate the component of the spins.
$J_{\xb\yb}$ is the isotropic coupling between sites $\xb$ and
$\yb$. $D_{\xb\yb}$ is the anisotropy operator: a $3\times3$ symmetric
matrix, where the six matrix elements $D_{\xb\yb}^{\alpha\beta}\,,
\alpha\geq\beta,$ are independent random variables.

There is quenched disorder, this means that the time scales of the
couplings $\{J_{\xb\yb},D_{\xb\yb}\}$ are infinitely larger than those of
our dynamic variables, so we represent them as constant in time random
variables, with $\overline{J_{\xb\yb}}=\overline{D_{\xb\yb}^{\alpha\beta}}=0$,
$\overline{J_{\xb\yb}^2}=1$ and $\overline{(D_{\xb\yb}^{\alpha\beta})^2}=D^2$
The overline $\overline{\cdots}$ denotes the averages over the instances of
the disorder, while for thermal averages we will use $\langle\cdots\rangle$.
Each different realization of the couplings $\{J_{\xb\yb},D_{\xb\yb}\}$ is
called \emph{sample}. Independent systems with the same couplings are
\emph{replicas} of the same sample. We use two replicas per sample.  

Notice that if all the matrix elements $D^{\alpha\beta}_{\xb\yb}$ are zero we
recover the fully isotropic Heisenberg model, with $O(3)$ symmetry. However,
if the $D^{\alpha\beta}_{\xb\yb}$ are non-vanishing, the only remaining
symmetry is time-reversal: $\vec s_\xb \longrightarrow -\vec s_\xb$ for all
the spins in the lattice. Time reversal is an instance of the $Z_2$
symmetry. This is the symmetry group of the IEA model.\cite{EA75} Hence, we
expect that the $Z_2$ symmetry will be spontaneously broken in a unique phase
transition belonging to the IEA Universality class (see
e.g. Ref.~\onlinecite{Gingras93}). Of course,
underlying this expectation is the assumption that the anisotropic coupling is
a relevant perturbation in the Renormalization Group sense (as it is the case
in ferromagnets~\cite{Amit05}). In fact, the infinite-anisotropy limit
can be explicitly worked out for a problem with \emph{site} anisotropy
[rather than link anisotropy as in Eq.~\eqref{eq:HamiltonianDefinition}]:
one finds an IEA-like behavior.\cite{Parisen06,Liers07}

It is widely accepted that the Universality class does not change with the
probability distribution of the couplings.\cite{Independence} We take
advantage of this, and choose a bimodal distribution for $J_{\xb\yb}$ and
$D_{\xb\yb}^{\alpha\beta}$, $J_{\xb\yb}=\pm 1$ and
$D_{\xb\yb}^{\alpha\beta}=\pm D$. These couplings can be stored in a single
bit, which is important because we are using GPUs, special hardware devices
where memory read/write should be minimized
(Appendix~\ref{sec:numericaldetails}).

We  chose the  two different  values $D  = 0.5,  1$.  We  want to  compare our
results with those  in Ref. \onlinecite{MartinPerez11}, where simulations
were  done  on  samples with  weak  random  anisotropies.   In that  work  the
$D_{\xb\yb}^{\alpha\beta}$  did not  follow a  bimodal distribution,  but were
uniformly distributed  between $-0.05$ and $0.05$. To  make proper comparisons
we  consider   the  standard  deviation  of  the distribution.  For bimodal
distributions  it is  exactly $D$,  in Ref. \onlinecite{MartinPerez11}  it is
$\overline{(D^2)^{1/2}}=1/\sqrt{1200}\simeq0.03$.

\subsection{The Observables}
\label{sec:obs}
To define the SG and CG order parameters we use two replicas. The overlap field is $q_\xb = \vec s^{\,a}_\xb\cdot \vec s^{\,b}_\xb$, 
where $a$ and $b$ are replica indices. Its Fourier Transform at wave vector
$\vn{k}$ is $\hat q_\mathrm{SG}(\vn{k}) = \sum_\xb q_\xb {\mathrm e}^{i \vn{k}\cdot \xb}/N$.

The chirality represents the oriented volume of the parallelepiped we can construct on 3 consecutive spins:
\begin{equation}
  \zeta_{\xb,\mu} = \vec s_{\xb+\vn{e}_\mu} \cdot (\vec s_\xb \times \vec s_{\xb-\vn{e}_\mu})  ~~~,~~~\mu=1,2,3,
\end{equation}
where $e_\mu$ is the unitary vector in the $\mu$ direction. 
The CG overlap is defined similarly to the SG one, as $\kappa_{\xb,\mu} =
\zeta_{\xb,\mu}^a \zeta_{\xb,\mu}^b$.  Again $a$ and $b$ indicate the
replica. The Fourier Transform of the CG overlap field is $\hat
q_\mathrm{CG}^\mu(\vn{k}) = \sum_\xb \kappa_\xb e^{i \vn{k}\cdot \xb}/N$.

We define the wave-vector dependent susceptibilities on the two overlap fields
as
\begin{equation}
  \chi_\mathrm{SG} = N \overline{\langle|q_\mathrm{SG}(\vn{k})|^2\rangle}~~,~~\chi_\mathrm{CG} = N \overline{\langle|q_\mathrm{CG}(\vn{k})|^2\rangle},
\end{equation}
and from each of them we can compute the correlation length of the related field \cite{Amit05}
\begin{equation}\label{eq:xi-second-moment}
  \xi = \frac{1}{2 \sin (k_\mathrm{min}/2)} \sqrt{\frac{\chi(0)}{\chi(\vn{k}_\mathrm{min})} -1},
\end{equation}
being ${\vn{k}_\mathrm{min}}=(2\pi/L,0,0)$ or permutations.
When computing $\xi_\mathrm{CG}$, one can choose $\mu$ parallel or orthogonal to the wave
vector $\vn{k}_\mathrm{min}$. As it was already observed in Ref. \onlinecite{Fernandez09}, there is no apparent difference between
the two options, so we averaged over all the values of $\mu$ to enhance
our statistics.

\section{Simulation details and Equilibration}
\label{sec:SimulationDetails}
We used Monte Carlo dynamics throughout all the work. Previous experience
advises to mix several Monte Carlo dynamics.\cite{JANUS12,Campos06,Lee07} In
fact, our single Monte Carlo step (MCS) consisted of (in successive order):
(i) one full lattice sweep with the heat-bath algorithm, (ii) $L$ lattice
sweeps of microcanonical overrelaxation algorithm,\cite{Brown87} and (iii) one
single Parallel Tempering sweep.\cite{Hukushima96,Marinari98} The combination
of the first two, which update one spin at a time, has been shown to be
effective in the case of isotropic SGs \cite{Pixley08} and other models with
frustration. \cite{Alonso96,Marinari00} Both heat-bath and overrelaxation are
directly generalized to the anisotropic case.\cite{generalize}

All the simulations were run on NVIDIA Tesla GPUs.  Except $L=64$, $D=0.5$,
where we parallelized 45 GPUs, each sample was simulated on a single GPU.  The
interested reader can find in appendix \ref{sec:numericaldetails} details on
how they were performed.

Table~\ref{tab:nsamples} depicts the relevant simulation parameters. For given
$L$ and $D$, the simulations were all equally long, except for $L=64$, $D=0.5$, where
we extended the simulation of the samples with the longest relaxation times.
\begin{table}
  \begin{ruledtabular}
    \begin{tabular}{ccccccc}
      $D$ & $L$ & $N_\mathrm{samples}$ & $N_\mathrm{MCS}^{\mathrm{min}}$& $N_\mathrm{T}$  & $T_\mathrm{min}$ & $T_\mathrm{max}$\\\hline\hline
      0.5 & 8   & 377                  & 2.048$\times10^4$&  10 & 0.588 & 0.8\\
      0.5 & 16  & 377                  & 4.096$\times10^4$ &  28 & 0.588 & 0.8\\
      0.5 & 32  & 377                  & 3.28$\times10^5$&  45 & 0.583 & 0.8\\
      0.5 & 64  & 185                  & 4$\times10^5$ &  45 & 0.621 & 0.709\\
\hline
      1   &  8  & 1024                 & 2.048$\times10^4$ &  10 & 0.877 & 1.28 \\
      1   & 12  & 716                  & 1.68$\times10^5$ &  20 & 0.893 & 1.28\\
      1   & 16  & 1024                 & 4.096$\times10^4$ &  28 & 0.877 & 1.28 \\
      1   & 24  & 716                  & 1.68$\times10^5$&  40 & 0.900 & 1.28 \\
      1   & 32  & 1024                 & 3.28$\times10^5$&  45 & 0.917 & 1.28\\
\hline
      1   & 64  & 54                   & 3.44$\times10^5$&  45 & 1.0 & 1.16009
    \end{tabular}
  \end{ruledtabular}
  \caption{Details of the simulations. We show the simulation
    parameters for each anisotropy $D$, and lattice size $L$.
    $N_\mathrm{samples}$ is the number of simulated samples.
    $N_\mathrm{T}$ is the number of temperatures that were used in
    parallel tempering. The temperatures followed a geometric sequence
    between $T_\mathrm{min}$ and $T_\mathrm{max}$, and $N_\mathrm{T}$
    was chosen so that the parallel tempering's acceptance was around
    $15\%$.  $N_\mathrm{MCS}^{\mathrm{min}}$ is the minimum number of
    MCS for each simulation. The simulation for $L=64$, $D=1$ was
    intended only to locate $T_\mathrm{CG}$.
}
  \label{tab:nsamples}
\end{table}

To ensure thermalization we made a \emph{logarithmic data binning}. Each bin
had twice the length of the previous, i.e. it contained two times more Monte
Carlo Steps (MCS), and had twice the measures. More explicitly, let us call
$i_\mathrm{f}$ the last bin: $i_\mathrm{f}$ contains the last half of the
Monte Carlo time series, $i_\mathrm{f}-1$ the second quarter, $i_\mathrm{f}-2$
the second octave, and so on.  This allowed us to create a sequence of values
$\langle O_n(i)\rangle$, for every observable $O$, where $n$ indicates the
sample, and $i$ identifies the bin, that has length $2^i$ MCS.  A set of
samples was considered thermalized if $\overline{\langle O_n(i)\rangle-\langle
  O_n(i_\mathrm{f})\rangle}$ converged to zero.  This test is stricter than
merely requesting the convergence of the sequence of $\overline{\langle
  O_n(i)\rangle}$, because neighboring blocks are statistically correlated, so
the fluctuation of their difference is smaller.\cite{Fernandez08} Physical
results were taken only from the last block.

Since the $L=64, D=0.5$ samples were the most GPU-consuming, we were more strict with
them.  To ensure and monitor thermalization, beyond the previous criteria, we
measured the integrated autocorrelation time (mixing time) of the random walk
in temperatures of each sample.\cite{Fernandez09} In a thermalized sample, all
the replicas stay a significant amount of time at each temperature. We made
sure that all the simulations were longer than 10 times this autocorrelation
time. The sample-to-sample fluctuations were not extreme, and the
autocorrelation times $\tau$ spanned between 10000 MCS to 50000 MCS, depending
on the sample.  Finally, we decided to take measures only over the last 64000
MCS of each simulation.

\section{Finite-Size Scaling}
\label{sec:FSS}
Our simulations were far from the thermodynamic limit, therefore in our
analysis we had to take in account finite-size effects.  Finite-Size Scaling
(FSS) consists in comparing results at different lattice size to characterize
the critical point. Specifically, we shall be employing
phenomenological-renormalization, also known as the quotients method.\cite{Nightingale76,Ballesteros96,Amit05}

Since FSS applies irrespectively of the considered order parameter, in the
current section we will not distinguish between spin and chiral sector.  The
generic critical temperature will be called $T_\mathrm{c}$.

If an observable $O$ diverges at the critical temperature as
$O\propto|T-T_\mathrm{c}|^{x_O}$, then its thermal average close to the
critical point can be expressed like
\begin{eqnarray}
  \langle O(L,T)\rangle &=& L^{x_O/\nu} \bigg [
    f_O(L^{1/\nu}\big(T-T_\mathrm{c})\big) \\\nonumber
    &+&             L^{-\omega}   g_O\big(L^{1/\nu}(T-T_\mathrm{c})\big)\\\nonumber
    &+&             L^{-2\omega}  h_O\big(L^{1/\nu}(T-T_\mathrm{c})\big) + \ldots \bigg],
  \label{eq:scalingGenerico}
\end{eqnarray}
where $f_O,g_O$ and $h_O$ are analytic scaling functions for observable $O$,
while $\nu$ is the thermal critical exponent.  The exponent $\omega>0$ is
universal, and it expresses the corrections to scaling. The lower dots stand
for sub-leading corrections to scaling.  Let us name $\xi_L(T)$ the
correlation length in a lattice of finite size $L$, at temperature $T$.  The
case $O=\xi_L(T)/L$ is of special interest, since $\nu$ is the critical
exponent for the correlation length. Then, Eq.~\eqref{eq:scalingGenerico}
becomes in this case, up to the leading-order,
\begin{equation}
  \frac{\xi_L}{L} = f_\xi\big(L^{1/\nu}(T-T_\mathrm{c})\big) +\ldots.
  \label{eq:xi}
\end{equation}
Therefore, we can identify $T_\mathrm{c}$ as the temperature where the curves $\xi_L(T)/L$ cross for all $L$ for sufficiently large $L$.
If we let $T^{L,2L}$ be the temperature where $\xi_L(T)/L$ crosses $\xi_{2L}(T)/(2L)$, this regime is
reached once the $T^{L,2L}$ has converged.
Yet, if $\omega$ is small, our lattice sizes may not be large enough, so we will have to take in account the aforementioned corrections to scaling.
Including corrections to the order $L^{-2\omega}$, the approach of the crossing temperature $T^{L,2L}$ to the asymptotic value $T_\mathrm{c}$
can be written as
\begin{equation}
  T^{L,2L}-T_\mathrm{c} = A L^{-(\omega+1/\nu)} + B L^{-(2\omega+1/\nu)} + \ldots\,,
  \label{eq:scalingCorrections}
\end{equation}
where $A$ and $B$ are non-universal scaling amplitudes.

To compute the critical exponents $\nu$ and $\eta$ we use the quotients' method, taking the quotient of the same observable between different lattice 
sizes $L$ and $2L$. At the temperature $T^{L,2L}$ we get:
\begin{equation}
  \frac{\overline{\langle O_{2L}(T^{L,2L})\rangle_J}}{\overline{\langle
      O_{L}(T^{L,2L})\rangle_J}} = 2^{x_O/\nu}+A_{x_O} L^{-\omega}+\ldots\,.
\end{equation}
Again, $A_{x_O}$ is a non-universal amplitude, while the dots stand for
subleading corrections to scaling.  Therefore, if $O$ is the thermal
derivative of $\xi$, we can compute the $\nu$ critical exponent through the
relation
\begin{equation}
  \frac{d\xi_{2L}(T^{L,2L})/dT}{d\xi_{L}(T^{L,2L})/dT} = 2^{1+1/\nu}+A_\nu L^{-\omega}+\ldots\,.
  \label{eq:nu}
\end{equation}
To calculate $\eta$ we use the susceptibility, as $\chi\propto|T-T_\mathrm{c}|^{-\gamma}$ and $2-\eta = \gamma/\nu$, hence
\begin{equation}
  \frac{\chi_{2L}(T^{L,2L})}{\chi_{L}(T^{L,2L})} = 2^{2-\eta}+A_\eta L^{-\omega}+\ldots\,.
  \label{eq:eta}
\end{equation}
Note that the value of $\xi_L/L$ at the crossing tends as well to a Universal
quantity:
\begin{equation}\label{eq:xi-L-cross}
  \left.\frac{\xi_L}{L}\right|_{T^{L,2L}}=
  \left.\frac{\xi^*}{L}\right|_{L=\infty}+A_\xi L^{-\omega}+\ldots\,.
\end{equation}

\section{Interpolations, extrapolations and errors}
\label{sec:methodology}
We have been able to estimate the critical temperature from the crossing of
the curves $\xi/L$ at $L$ and $2L$, and the exponents $\nu$ and $\eta$ with
the method of the quotients, as described in section \ref{sec:FSS}.

To identify the crossing point between the pairs of curves, we used low-order
polynomial fits: for each lattice size, we took the four temperatures in the
parallel tempering nearest to the crossing point. We fitted these four data
points to a linear or quadratic function of the temperature.  The obtained
results were compatible within one standard deviation (the values reported in
this work come from the linear interpolation).  In order to calculate $\nu$ we
needed the derivative of the correlation length at the crossing point. We
extracted it by taking the derivative of the polynomial interpolations.

However, there is a difficulty in the calculation of statistical errors: the
fits we had to perform came from strongly correlated data (because of the
parallel-tempering temperature swap). Therefore, to get a proper estimate of
the error, we made Jack-knife blocks, fitted separately each block, and
calculated the Jack-knife error.\cite{Amit05}

The whole mentioned procedure was fluid while $T_\mathrm{SG}^{L,2L}$ fell in
our simulated temperature span.  Yet, since $T_\mathrm{SG}^{L,2L}$ was fairly
lower than $T_\mathrm{CG}^{L,2L}$, it occurred in four cases that we did not
reach low enough temperatures in our simulations to be able to interpolate the
crossing, and we had to recur to extrapolations.  This happened with $D=1$,
$T_\mathrm{SG}^{32,64}$ and $T_\mathrm{CG}^{32,64}$, and in
the lower
anisotropy $D=0.5$, with $T_\mathrm{SG}^{16,32}$ and $T_\mathrm{SG}^{32,64}$.

The case of $T_\mathrm{SG}^{32,64}(D=1)$ and $T_\mathrm{SG}^{16,32}(D=0.5)$
was not a great issue, because the crossing point was very
near to the lowest simulated temperature, so we treated these crossings just
like the others.  

In the case of $T_\mathrm{SG}^{32,64}(D=0.5)$, instead, we had to
extrapolate at a long distance (see Fig.~\ref{fig:xiL_SG}--top, in the next
section). Again, we performed the extrapolation through linear in temperature
fits. To make the fit of $L=64$ more stable, we took in account a progressive
number of points (i.e. we fitted to the $n$ lowest temperatures). We increased
the number of temperatures, while the crossing temperature was constant. Note
that increasing the number of temperatures in the fit results in a smaller
statistical error for the crossing-temperature. However, $\xi_L(T)/L$ is not a
linear function at high $T$ (see Fig.~\ref{fig:xiL_SG}). Therefore a tradeoff
is needed because, when too
high temperatures were included in the fit, the crossing temperature started
to change, and we knew that curvature effects were biasing it. Our final
extrapolation was obtained from a fit performed on the 10 lowest-temperature
points. Unfortunately, this approach was not feasible for the SG
susceptibility due to its strongly non-linear behavior. Hence, in the next
section we will not give an estimate for $\eta_\mathrm{SG}(L=64)$.

In the case of $T_\mathrm{SG}^{32,64}(D=1)$, the simulation was not
devised to reach that crossing point, and we did not extrapolate data.

\section{Results}
\label{sec:results}

\subsection{Spin Glass Transition}
\begin{figure}[b]
  \includegraphics[angle=270, width=\columnwidth]{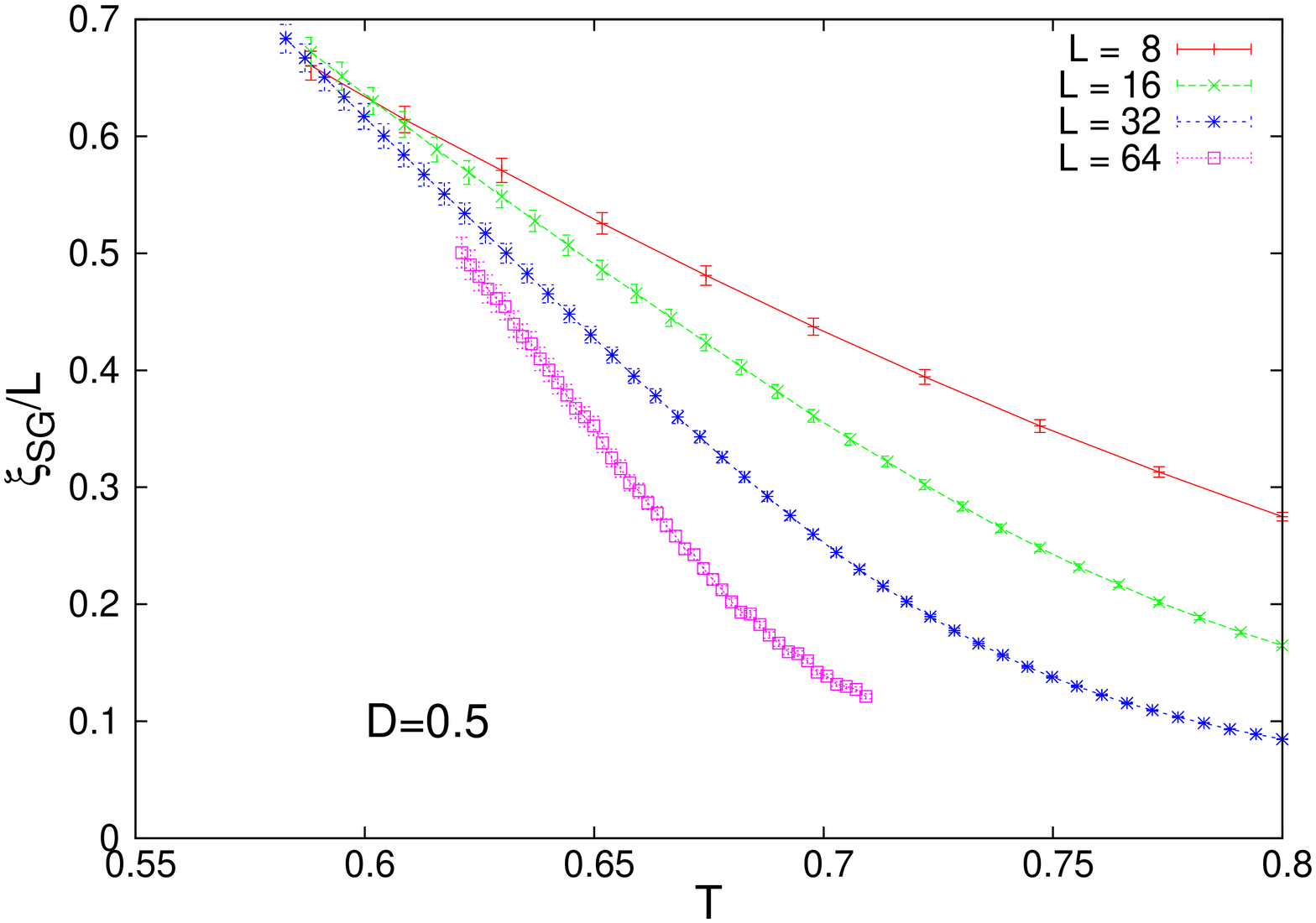}
  \includegraphics[angle=270, width=\columnwidth]{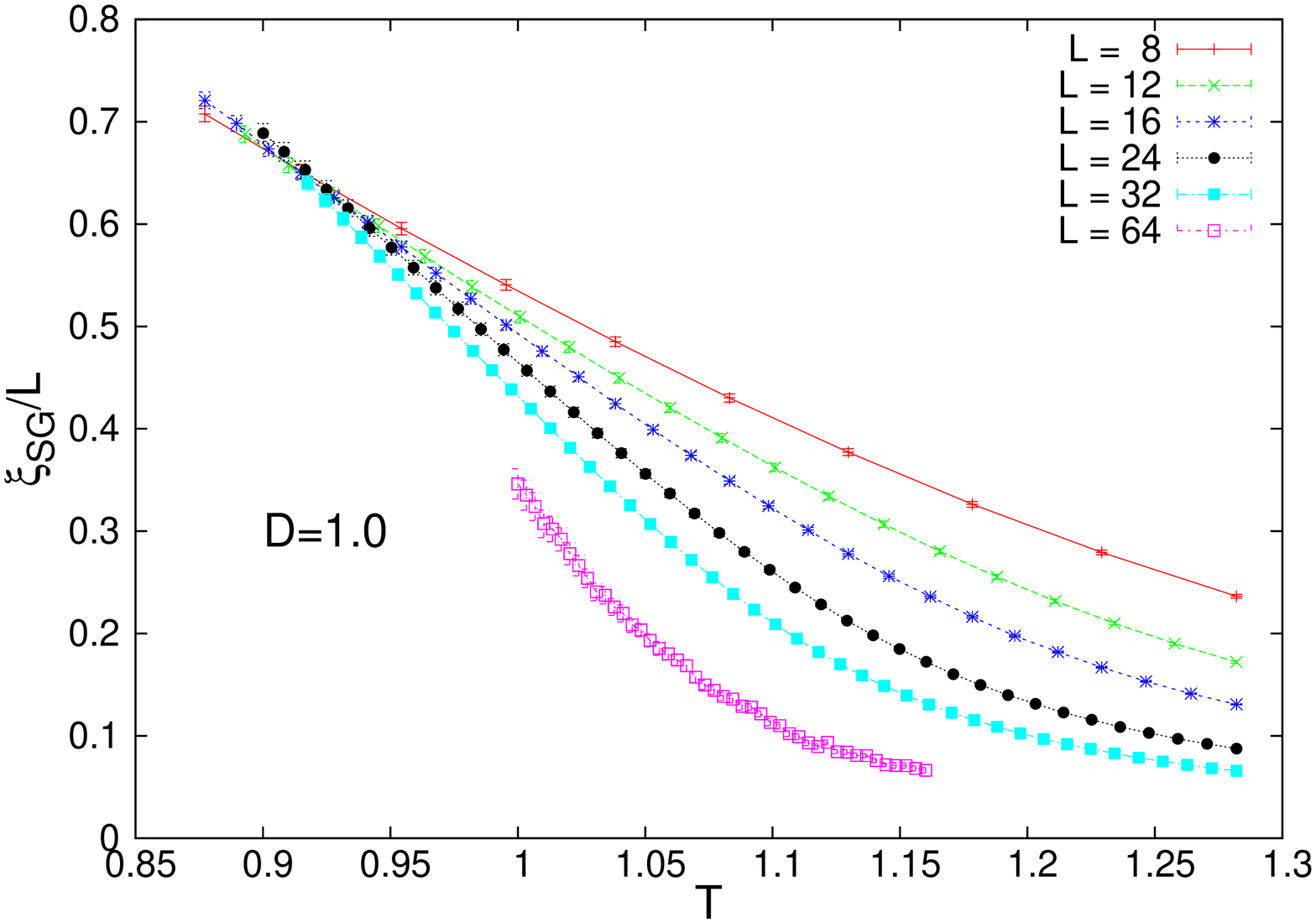}
  \caption{(color online). Spin glass correlation length in units of
    the linear lattice size $L$ for $D=0.5$ (\textbf{top}) and $D=1$
    (\textbf{bottom}).  All the curves cross at about the same
    temperature for both anisotropies (see
    Eq. \ref{eq:scalingCorrections}). The data for $D=1$, $L=64$,
    shown here for the sake of completeness, were only used for the
    chiral sector.}
  \label{fig:xiL_SG}
\end{figure}

Figures \ref{fig:xiL_SG} show the crossings of $\xi_\mathrm{SG}(T)/L$ for $D=0.5, 1$. Table \ref{tab:TSG} contains the principal results on the SG
sector, providing a quantitative description of those figures.
As explained in Sect.~\ref{sec:model-symmetries}, we expect that the
transition belongs to the Ising-Edwards-Anderson (IEA) Universality class.
This conjecture is supported by the fact that the critical exponents
$\nu_\mathrm{SG}$ and $\eta_\mathrm{SG}$, and the height at which the
$\xi_\mathrm{SG}(T)/L$ cross, are compatible with those of the IEA spin glass,
indicated in the last line of table \ref{tab:TSG}.

\begin{table}[ht]\centering
  \begin{center} 
    \emph{Determination of the critical quantities for the SG sector.}
  \end{center}
  \begin{ruledtabular}
    \begin{tabular}{cccccc}
      $D$ & $(L,2L)$ & $T_\mathrm{SG}$   & $\nu_\mathrm{SG}$ & $\eta_\mathrm{SG}$ & $\xi_\mathrm{SG}(T_\mathrm{SG})/L$\\\hline\hline
      0.5 & (8,16)     & 0.602(18)  & 1.91(27)   & -0.388(27)  & 0.629(48)\\
      0.5 & (16,32)    & 0.577(22)  & 2.70(63)   & -0.449(67)  & 0.705(76)\\
      0.5 & (32,64)    & 0.596(14)  & 2.18(45)   & -           & 0.631(56)\\\hline
      0.5 & $\infty$ & \footnotesize 0.591(16)[0]  & \footnotesize 2.71(82)[3]   & -           & \footnotesize 0.637(87)[1]\\
      \multicolumn{2}{c}{$\chi^2/\mathrm{d.o.f.}$} & 0.55/1   &  0.47/1  & -           & 0.56/1\\\hline\hline
      1.0 & (8,16)     & 0.910(21)  & 2.38(25)   & -0.410(44)  & 0.660(34)\\
      1.0 & (12,24)    & 0.927(19)  & 2.32(28)   & -0.370(53)  & 0.629(36)\\
      1.0 & (16,32)    & 0.910(16)  & 2.37(28)   & -0.400(19)  & 0.660(35)\\\hline
      1.0 & $\infty$ & \footnotesize 0.917(32)[0] & \footnotesize 2.33(67)[0]  & \footnotesize -0.391(71)[1] & \footnotesize 0.662(83)[0]\\
      \multicolumn{2}{c}{$\chi^2/\mathrm{d.o.f.}$}  &  0.66/1 &  0.030/1   & 0.37/1         & 0.55/1\\\hline\hline
      IEA  & $\infty$ &             & 2.45(15)   & -0.375(10)  & 0.645(15)
    \end{tabular}        
  \end{ruledtabular}
  \caption{\label{tab:TSG} For each anisotropy $D$, and each pair of
    lattices $(L,2L)$, we obtain effective size-dependent estimates
    for $T_\mathrm{SG}$, and the universal quantities
    $\nu_\mathrm{SG}$, $\eta_\mathrm{SG}$ and
    $\xi_L(T_\mathrm{SG})/L$. The thermodynamic limit, indicated with
    $L=\infty$, is obtained by means of fits to
    equations~\eqref{eq:scalingCorrections},~\eqref{eq:nu},~\eqref{eq:eta}
    and~\eqref{eq:xi-L-cross}. Exponent $\omega$ was not a fitting
    parameter (we took $\omega_\mathrm{IEA}=1.0(1)$ from
    Ref. \onlinecite{Hasenbusch08}, see text and endnote
    \onlinecite{janus-compatible}).  The line immediately after the
    extrapolations displays the estimator of the $\chi^2$ figure of
    merit of each one.  $D=\mathrm{IEA}$ represents the critical
    values of the Ising-Edwards-Anderson Universality class, taken
    from Ref. \onlinecite{Hasenbusch08}.  The numbers in square
    brackets express the systematic error due to the uncertainty of
    $\omega_\mathrm{IEA}$.  }
\end{table}

Hence, it is reasonable to extrapolate our results to $L\rightarrow\infty$ by
assuming the IEA Universality class. We took $\omega_\mathrm{IEA}=1.0(1)$ from
Ref. \onlinecite{Hasenbusch08}, and fitted to
Eqs.~\eqref{eq:nu},~\eqref{eq:eta} and~\eqref{eq:xi-L-cross}.  In those fits
we took in account both the anticorrelation in the data,\cite{covariance} and
the bias arising from the indetermination of the exponent
$\omega_\mathrm{IEA}$.  Notice, from table \ref{tab:TSG}, that the dependence
on $L$ of the data is so weak, that this bias is practically negligible.
This situation is different from the one encountered in
Ref. \onlinecite{MartinPerez11}, where the anisotropy fields were extremely
small ($D\simeq0.03$). \cite{Dequivalence} There, the finite-size effects in
the SG sector were huge.

Overall, the strong consistency of our extrapolations to large $L$ with the
IEA exponents shows \emph{a posteriori} that our assumption was proper.

\subsection{Chiral Glass Transition}
\begin{figure}[htb]
  \includegraphics[angle=270, width=\columnwidth]{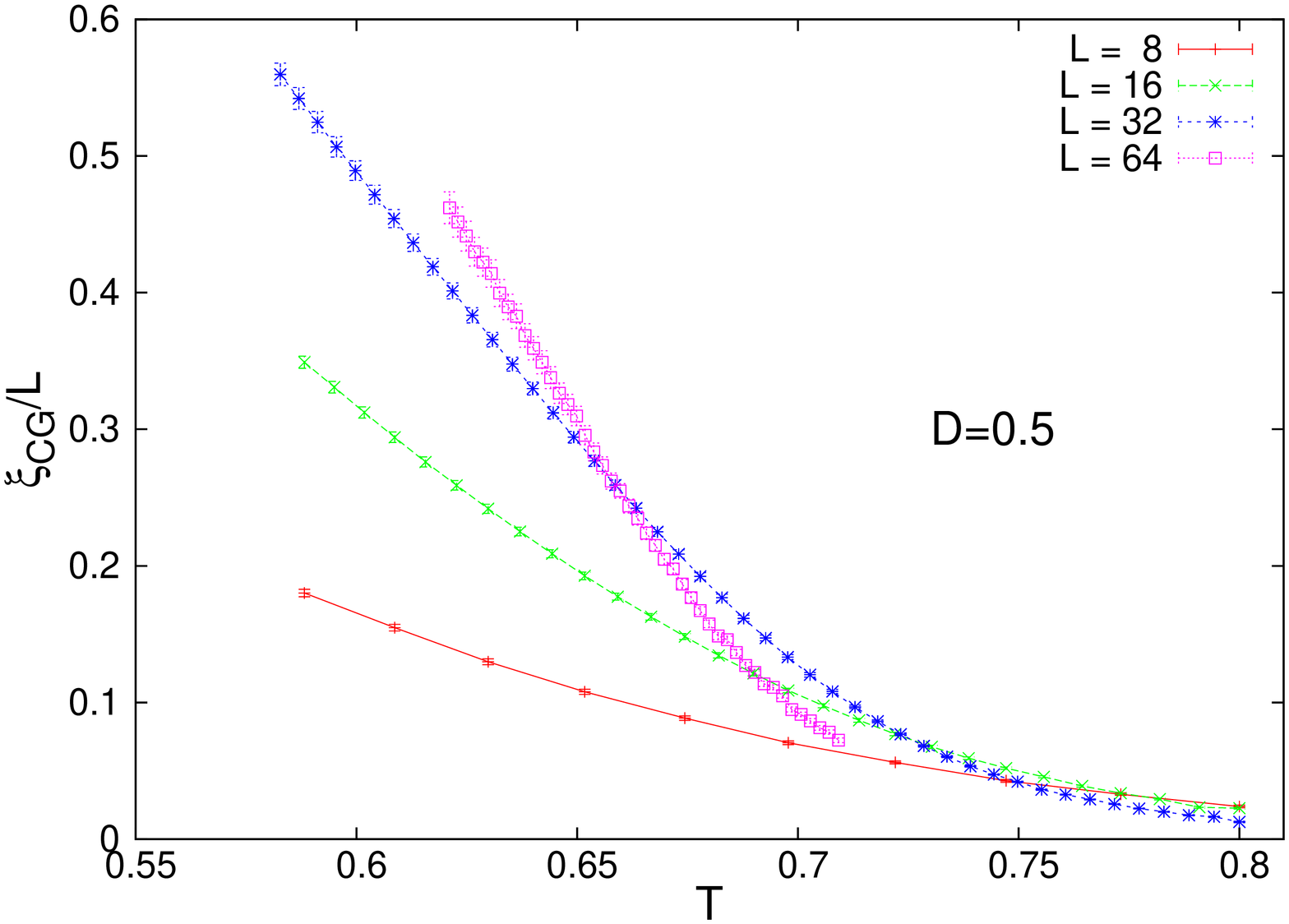}
  \includegraphics[angle=270, width=\columnwidth]{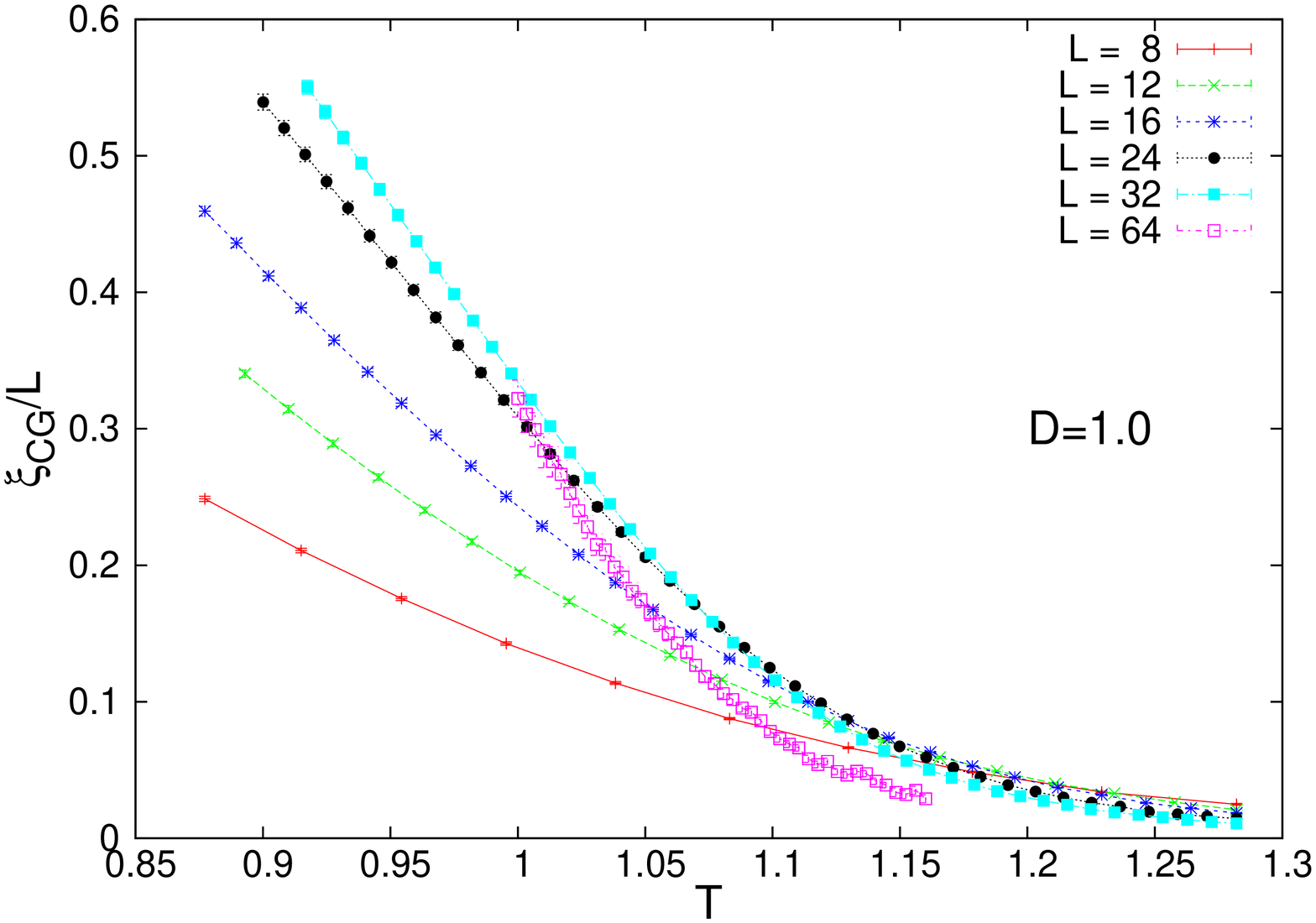}
  \caption{(color online). Chiral Glass correlation length in units of the
    lattice size for $D=0.5$ (\textbf{top}) and $D=1$ (\textbf{bottom}). When
    $L$ grows, the crossing temperature shifts significantly towards left.}
  \label{fig:xiL_CG}
\end{figure}

In the CG channel (figures \ref{fig:xiL_CG} and table \ref{tab:TCG}) the interpretation is slightly more controversial, since 
finite-size effects are heavy.
For the smaller lattice sizes, $T_\mathrm{CG}$ is consistently larger than $T_\mathrm{SG}$, and $\nu_\mathrm{CG}$ is incompatible
with the IEA limit. On the other side, when $L$ is larger, $T_\mathrm{CG}$ 
approaches noticeably its SG counterpart, and so does $\nu_\mathrm{CG}$.
We notice that $\eta_\mathrm{CG}$ marks the distinction between these two
regimes. In fact, when $L$ is small, it is very close to 2.  This means that
the divergence of $\chi_\mathrm{CG}$ is extremely slow ($\chi\sim
L^{2-\eta}$),\cite{scalingRelation} revealing we are still far from the
asymptotic limit. When $L$ is larger, $\eta_\mathrm{CG}$ is consistently
smaller, the divergence of $\chi_\mathrm{CG}$ is less suppressed, and we can
assume the asymptotic behavior is starting to show up. Consistently with this
observation, the value of $\xi_\mathrm{CG}/L$ at the crossing temperature
becomes sizeable [indeed, the second-moment correlation length
  \eqref{eq:xi-second-moment} is well defined only if $\eta<2$, see
  e.g. Ref.~\onlinecite{Amit05}].
\begin{table}[ht]
  \begin{center} 
    \emph{Determination of the critical quantities for the CG sector.}
  \end{center}
  \begin{ruledtabular}
    \begin{tabular}{cccccc}
      $D$ & $(L,2L)$ & $T_\mathrm{CG}$ & $\nu_\mathrm{CG}$ & $\eta_\mathrm{CG}$ & $\xi_\mathrm{CG}(T_\mathrm{CG})/L$\\\hline\hline
      0.5 &  (8,16)    & 0.7762(43) & 1.45(22)   & 1.9778(23)  & 0.0321(22)\\
      0.5 &  (16,32)   & 0.7255(29) & 1.78(14)   & 1.8416(98)  & 0.0735(41)\\
      0.5 &  (32,64)   & 0.659(47)  & 2.40(47)   & 0.823(68)   & 0.258(18)\\\hline\hline
      1.0 &  (8,16)    & 1.2031(33) & 1.205(71)  & 1.9507(27)  & 0.0418(12)\\
      1.0 &  (12,24)   & 1.1472(40) & 1.72(11)   & 1.8664(51)  & 0.0691(25)\\
      1.0 &  (16,32)   & 1.1046(38) & 2.18(10)   & 1.6995(75)  & 0.1098(42)\\
      1.0 &  (32,64)   & 0.987(22)  & 2.48(84)   & 0.53(19)    & 0.368(58)
    \end{tabular}
  \end{ruledtabular}
  \caption{Same as table \ref{tab:TSG}, but for chirality. In
    this case the corrections to scaling are significant.}
  \label{tab:TCG}
\end{table}

\subsection{Uniqueness of the transition}
Although the SG and CG transitions do not coincide yet with our values of $L$ and $D$, the 
critical temperatures, as well as $\nu$,
become more and more similar as the linear size of the system increases. 
Moreover, the decrease of $\eta_\mathrm{CG}$ as a function of $L$ has not yet stabilized, 
so it is likely that the chiral quantities
will keep changing with bigger lattice sizes.

\begin{figure}[b]
  \includegraphics[angle=0, width=\columnwidth]{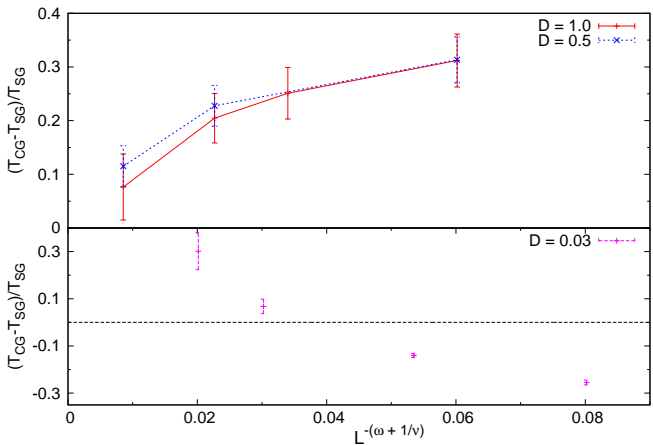}
  \caption{(color online). Difference between the chiral crossing $T_\mathrm{CG}$ 
    and the spin glass transition temperature $T_\mathrm{SG}^\infty$, 
    in units of $T_\mathrm{SG}^\infty$ (see Table~\ref{tab:TSG} 
    for the extrapolations of $T_\mathrm{SG}^\infty$).  The exponents
    $\omega_\mathrm{IEA}$ and $\nu_\mathrm{IEA}$ are taken from
    Ref. \onlinecite{Hasenbusch08}. 
    In the \textbf{upper plot} we represent our
    data, for $D=0.5,1$.  The two transitions get closer when we increase $L$,
    and the approach appears faster when the lattice size increases. Notice that
    a linear interpolation between the
    two largest lattice sizes intercepts the $y$ axis compatibly with a coupling
    between the two transitions (i.e. $T_\mathrm{SG}=T_\mathrm{CG}$).  On the
    \textbf{bottom plot} we show data from Ref. \onlinecite{MartinPerez11},
    where much lower anisotropies were considered. Here the scenario is
    completely different, since the critical temperatures drift apart for large
    enough $L$. The horizontal dashed line corresponds to
    $T_\mathrm{CG}-T_\mathrm{SG}=0$.}
  \label{fig:deltaTc}
\end{figure}

As explained in Sect.~\ref{sec:model-symmetries}, we expect that the
transition should belong to the IEA Universality class.  To confirm this
expectation, we make the ansatz of a unique transition, of the IEA Universality
class, to seek if the two critical temperatures join for $L\rightarrow\infty$.
Figure \ref{fig:deltaTc} (upper half) shows the difference between the
critical temperatures as a function of the natural scale for first order
corrections to scaling, $L^{-(\omega_\mathrm{IEA}+1/\nu_\mathrm{IEA})}$
[Eq. \eqref{eq:scalingCorrections}].  Again, $\omega_\mathrm{IEA}$ and
$\nu_\mathrm{IEA}$ are taken from Ref. \onlinecite{Hasenbusch08}.  Not only
Fig. \ref{fig:deltaTc} (top) reveals a marked increase of the speed of the
convergence for $L=64$ (to which corresponds the smallest anomalous exponent
$\eta_\mathrm{CG}$), but also, a linear interpolation to infinite
volume, taking that point and the previous, extrapolates
$T_\mathrm{SG}=T_\mathrm{CG}$ within the error.

\begin{figure}[b]
  \includegraphics[angle=0, width=\columnwidth]{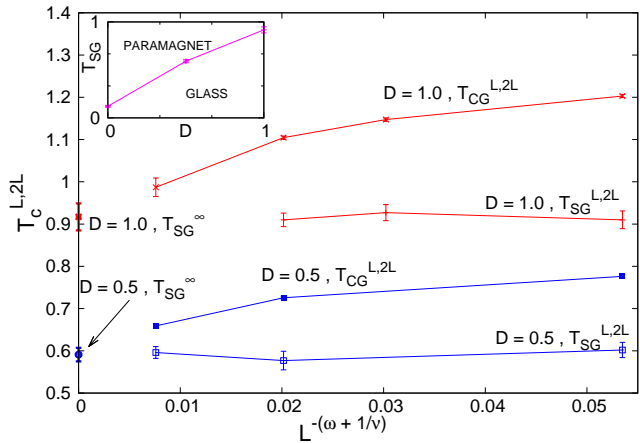}
  \caption{(color online). Crossing temperatures as a function of $L^{-(\omega_\mathrm{IEA}+1/\nu_\mathrm{IEA})}$ (\textbf{large plot}). The points on the intercept are the $L\rightarrow\infty$ extrapolations
    from table \ref{tab:TSG}. The \textbf{inset} shows the phase diagram of the model with these same points, as 
    the most economic interpretation of our data is that
    in the thermodynamic limit
    $T_\mathrm{SG}=T_\mathrm{CG}$. The $D=0$ point is borrowed from Ref. \onlinecite{Fernandez09}.}
  \label{fig:Tc_vs_L}
\end{figure}

Fig. \ref{fig:Tc_vs_L} shows how the SG and CG critical temperatures
approach each other with $L$. Again, $T_\mathrm{CG}$ gets closer to
$T_\mathrm{SG}$, and the speed of the approach increases with the lattice
size. 
The points in the intercept represent extrapolations to the
thermodynamic limit of the $T_\mathrm{SG}$. 
Since the observations are compatible with the ansatz of a unique phase transition,
belonging to the IEA universality class,
we used the infinite-size limit of $T_\mathrm{SG}$ to plot the
model's phase diagram (Fig. \ref{fig:Tc_vs_L}, inset).\cite{PDdisclaimer}

\subsection{Comparing with weak anisotropies}
Both plots of Fig. \ref{fig:deltaTc} show the same observable, for different
anisotropies.  The top plot depicts our data, in the case of strong
anisotropies $D=0.5,1$. The bottom one represents the case of weak
anisotropies ($D\simeq0.03$),\cite{Dequivalence} coming from
Ref. \onlinecite{MartinPerez11}.  The behavior is very different between the
two cases.  For strong anisotropies, the critical temperatures tend to meet as
we increase $L$.  That is qualitatively very different from the weak
anisotropy case, where their distance increases.  We can ask ourselves where
this qualitative difference of behavior comes from.

If we compare same system sizes and different $D$ in table \ref{tab:TCG}, 
we notice that finite-size effects are larger (and $\eta$ closer to two) 
the smaller the anisotropy. These differences in the finite-size effects 
are appreciable with a factor 2 change in the anisotropy (from $D=1$ to $D=0.5$), 
so it is reasonable that suppressing the anisotropy by a factor 17 or 35
will increase drastically the finite-size effects.

The most economic explanation 
is then that there is a non-asymptotic effect that disappears with much larger systems or, as we have 
seen, with larger anisotropies. In other words there is a $L^*(D)$ after which $T_\mathrm{SG}$ and $T_\mathrm{CG}$ start joining. For $D\simeq0.03$, $L^*$ is so large that we observe a
growing $T_\mathrm{CG}-T_\mathrm{SG}$, while for $D\geq0.5$ we find $L^*<8$.

Another peculiarity outcoming from Ref. \onlinecite{MartinPerez11} arises from
the SG transition alone.  It had been observed that a very weak perturbation
on the symmetry of the isotropic system implied huge changes in the critical
temperature, while one would expect that the transition line is smooth.

\begin{figure}[ht]
  \includegraphics[angle=0, width=\columnwidth]{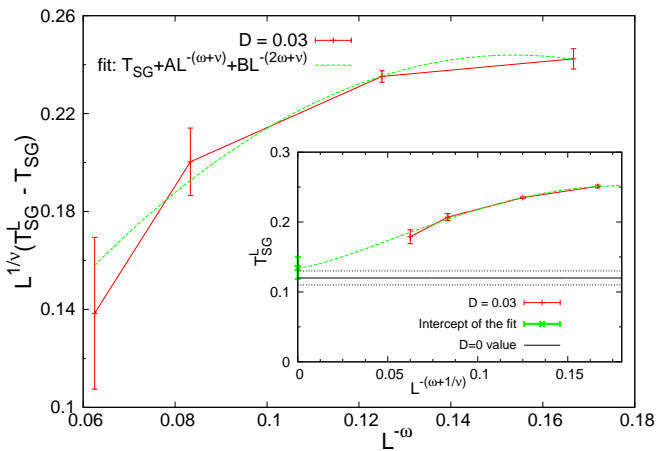}
  \caption{(color online). Data from \onlinecite{MartinPerez11}, corresponding
    to $D\simeq0.03$, \cite{Dequivalence} with extrapolations to the
    thermodynamic limit assuming the Ising-Edwards-Anderson Universality class.
    The data is the same in both plots. The dashed line is a fit of the scaling
    in $L$, considering corrections up to the second order
    (Eq. \ref{eq:scalingCorrections}).  The \textbf{large} figure displays the
    trend of the scaling variable $L^{1/\nu}(T-T_\mathrm{SG})$ as a function of
    $L^{-\omega}$. The \textbf{inset} shows the same data set, plotting
    $T_\mathrm{SG}^{L,2L}$ as a function of $L^{-\omega-1/\nu}$, see
    Eq.~\eqref{eq:scalingCorrections}. The extrapolation to large-$L$
    (the point in the intercept) is compared with $T_\mathrm{SG}$ of $D=0$
    from \onlinecite{Fernandez09}. The full horizontal line is the central value
    of $T_\mathrm{SG}^{D=0}$, and the dashed lines define the error.  }
  \label{fig:TSG_D003}
\end{figure}

To solve this dilemma, we take advantage of having strong evidence for the Universality class of the transition.
So, we take the data from Ref. \onlinecite{MartinPerez11}, and use once again the exponents $\nu_\mathrm{IEA}$ and $\omega_\mathrm{IEA}$ in 
Ref. \onlinecite{Hasenbusch08} to extrapolate the infinite volume limit
with second order corrections to scaling (Eq. \ref{eq:scalingCorrections}).
The fit is good ($\chi^2/\mathrm{d.o.f.}=0.70/1$), and, as we show in Fig. \ref{fig:TSG_D003}, its $L\rightarrow\infty$ extrapolation
for the critical temperature is 
compatible with $T_\mathrm{SG}(D=0)$ within one standard deviation.
Thus, taming the finite-size effects was enough to make the scenario consistent,
and the issue reduces to the fact that finite-size effects are extremely strong when 
the anisotropy is smaller.

\section{Conclusions}
\label{sec:conclusion}
We performed a numerical study of the critical behavior of Heisenberg Spin
Glasses with strong bimodal anisotropies. Our aim was to clarify the role of
scaling-corrections, as well as the crossover effects between the Heisenberg
and Ising Universality classes, to be expected when the anisotropic
interactions are present. In fact, we show that anisotropic interactions are a
relevant perturbation in the Renormalization Group sense: no matter how small 
the anisotropy, the asymptotic critical exponents are those of the Ising-Edwards-Anderson 
model. However, a fairly large correlation length maybe needed to reach the asymptotic regime.
This observation is relevant for the interpretation of both numerical
simulations,\cite{MartinPerez11} and experiments.\cite{Petit02}

It is then clear that large system sizes are needed to make progress,
something that calls for extraordinary simulation methods. Therefore, we
performed single-GPU and multi-GPU simulations to thermalize lattices up to
$L=64$ at low temperatures. As side benefit, our work provides a
proof-of-concept for GPU and multi-GPU massive simulation of spin-glasses with
continuous degrees of freedom. This topic is elaborated further in
Appendix~\ref{sec:numericaldetails}.

We performed a finite-size scaling analysis based on phenomenological
renormalization.\cite{Nightingale76,Ballesteros96} We imposed scale-invariance
on the second-moment correlation length in units of the system size,
$\xi_L/L$.  We followed this approach for both the chiral and spin glass order
parameters.

Our results for the spin-glass sector were crystal clear: all the indicators
of the Universality class were compatible with their counterparts in the
Ising-Edwards-Anderson model. On the other hand, in the chiral sector
scaling-corrections were annoyingly large, despite they decrease upon
increasing the magnitude of the anisotropic interactions.

Regarding the coupling of chiral and spin glass transition, our
numerical results seem to indicate that the two phase-transitions take
place at the same temperature
(i.e. $T_\mathrm{CG}=T_\mathrm{SG}$). However, it is important to
stress that we need our very largest lattices to observe this trend.
Nevertheless, what we see is in agreement with both Kawamura's
prediction and experiments, where the phase transitions are apparently
coupled, and the chiral glass susceptibility is
divergent.\cite{Taniguchi07}

Moreover, we were able to rationalize the
numerical results in Ref.~\onlinecite{MartinPerez11} with corrections to
scaling, by assuming the Ising-Edwards-Anderson Universality class.

We remark that there are strong analogies between the interpretation of
numerical and experimental data. In both cases, there is a relevant
length-scale (the correlation length for experiments, the system size for
simulations). If that length is large enough, the asymptotic
Ising-Edwards-Anderson Universality class should be observed. Otherwise,
intermediate results between Heisenberg and Ising are to be expected, and
indeed appear.\cite{Petit02} 

The difficulty in reaching the asymptotic regime lies on time: the time growth
of the correlation length is remarkably slow ($\xi(t_\mathrm{w})\sim
t_\mathrm{w}^{1/z}$ with $z\approx 7$,\cite{Belletti08,Joh99} where
$t_\mathrm{w}$ is the waiting time). Indeed, the current experimental record
is around $\xi\sim$ 100 lattice spacings,\cite{Joh99,Bert04} pretty far from
the thermodynamic limit.\cite{Ldiunamole} Hence attention should shift to the
study of the intermediate cross-over regime. An intriguing possibility
appears: one could envisage an experimental study of the crossover effects as
a function of the \emph{waiting time}. In fact, $t_w$ varies some four orders
of magnitude in current experiments,\cite{Rodriguez13} which should result in
a factor 4 variation of $\xi(t_\mathrm{w})$.

\subsection*{Acknowledgments}
We thank J.~J. Ruiz-Lorenzo and E. Marinari for useful discussions.  We were
partly supported by MINECO, Spain, through the research contract
No. FIS2012-35719-C02. MBJ was supported by the FPU program (MECD, Spain).
The research leading to these results has received funding from the European
Research Council under the European Union's Seventh Framework Programme
(FP7/2007-2013) / ERC grant agreement n° [247328].  The computations were
carried out in the GPU-accelerated clusters Tianhe-1A (Tianjin, China) and
Minotauro (Barcelona, Spain).  The total amount of time devoted to this
project was $ 2.2 \times 10^5$ GPU hours in Tianhe-1A and $2.0 \times 10^5$
GPU hours in Minotauro.  Access to Tianhe-1A was granted through research
contract No.  287746 by the EU-FP7. The authors thankfully acknowledge the
computer resources, technical expertise and assistance provided by the staff
at the {\em National Supercomputing Center-Tianjin} and at the {\em Red
  Espa\~nola de Supercomputaci\'on--Barcelona Supercomputing Center\/}.

\appendix

\section{Spin Glasses on (multiple) GPUs}
\label{sec:numericaldetails}

The appendix is structured as follows. The specific algorithms that we have
used are explained in Sect.~\ref{sec-app:algorithms} with no reference to
their implementation. However, implementation \emph{is} crucial: our
simulations are so demanding that we have used special hardware described in
Sect.~\ref{sec-app:hardware}. 
This special hardware speeds up the simulations
thanks to parallelization, so in Sect.~\ref{sec-app:parallelization} we give
some brief details about it.
Finally, we address in
Sect.~\ref{sec-app:PRNG} some issues regarding the generation of pseudo-random
numbers.

\subsection{Simulation algorithms}\label{sec-app:algorithms}

As explained in Sect.~\ref{sec:SimulationDetails}, we used a blend
of several Monte Carlo dynamics. Specifically, our single Monte Carlo step
(MCS) consisted of (in successive order):
\begin{itemize}
\item 1 full lattice sweep with the Heat-Bath algorithm, 
\item $L$ lattice sweeps of microcanonical overrelaxation algorithm,
\item 1 Parallel Tempering sweep.\cite{Hukushima96,Marinari98}
\end{itemize}
Heat-bath by itself would provide correct (but inefficient) dynamics. It
actually mimics the natural evolution followed by real spin glasses (that
never reach equilibrium near or below the critical temperature). For this
reason we enhance it with two more algorithms. However, heat-bath does play a
crucial role, since it is irreducible (i.e. the full configuration space is
reachable, at least in principle), at variance with overrelaxation, which
keeps the total energy constant, and parallel-tempering, which changes the
temperature but not the spin configuration.

Crucial to perform the heat-bath and overrelaxation dynamics is a
factorization property of the Boltzmann weight for the
Hamiltonian~\eqref{eq:HamiltonianDefinition}. The conditional
probability-density for spin $\vec s_{\xb}$, given the rest of the spins of
the lattice is
\begin{equation}\label{eq:ProbabilidadCondicionada}
  P(\vec s_{\xb}\,|\, \{\vec s_{\yb}\}_{\yb\neq\xb})\propto e^{(\vec s_{\xb}\cdot\vec h_{\xb})/T}\,,
\end{equation}
where $\vec h_{\xb}$ is the \emph{local field} produced by the lattice
nearest-neighbors of spin $\vec s_{\xb}$ (its precise definition is given in
footnote~\onlinecite{generalize}).

In the heat-bath update, a new orientation for spin $\vec s_{\xb}$ is drawn
from the conditional probability~\eqref{eq:ProbabilidadCondicionada}, see
Ref.~\onlinecite{Amit05} for instance.

The overrelaxation update is deterministic. Given a spin $\vec s_{\xb}$ and
its local field, we change the spin as much as possible while keeping the
energy constant:
\begin{equation}
  \vec s_{\xb}^\mathrm{\,new}=2\vec h_{\xb}\frac{\vec h_{\xb}\cdot\vec
    s_{\xb}^\mathrm{\,old}}{h_{\xb}^2} - \vec s_{\xb}^\mathrm{\;old}\,.
\end{equation}
Contrarily to heat-bath, the order in which the spins are updated is
important in overrelaxation. Accessing the lattice randomly increases the
autocorrelation time in a substantial way. On the other hand, a sequential
update generates a microcanonic wave that sweeps the lattice. The resulting
change in the configuration space is significantly larger. A similar
microcanonic wave is generated with other types of deterministic lattice
sweeps. For instance, one could partition the lattice in a checker-board way
and first update all spins in the black sublattice, updating the white spins
only afterwards.

The combination of heat-bath and overrelaxation has been shown to be effective
in the case of isotropic spin glasses \cite{Pixley08} and other models with
frustration. \cite{Alonso96,Marinari00} However, if one is interested on very
low temperatures or large systems, parallel tempering is often useful.  For
each sample we simulate $N_T$ different copies of the system, each of them at
one of the temperatures $T_1<T_2<\ldots<T_{N_T}$. A parallel tempering update
consists in proposing, as configuration change, a swap between configurations
at neighboring temperatures. The exchange has the Metropolis
acceptance. Evidently, the acceptance is higher if the temperatures $T_i$ are
closer to each other, since the energy of the configurations will be
similar. Notice that exchanging configurations is equivalent to exchange
temperatures, so the data transfer is reduced to a single number.

\subsection{Hardware features}\label{sec-app:hardware}

The GPUs we used were of the Tesla generation, produced by NVIDIA, with an
SIMD architecture (Single Instruction, Multiple Data), \cite{nvidia} optimized
for the parallel processing of large amounts of double precision data. 

We had
access to Tesla M2050 GPUs in the \emph{Tianhe-1A} supercomputer in Tianjin,
\cite{nscc} China, and Tesla M2090 GPUs on the \emph{Minotauro} cluster
\cite{bsc} in Barcelona, Spain. Despite the extremely high performances
claimed by NVIDIA (e.g. 665 Gflops in double precision in the case of the
M2090 GPUs), it is practically impossible to reach that limit, because the
major bottleneck does not reside in the computing speed, but in the memory
access.
Yet GPUs keep being a valid tool to simulate on spin glasses, as they
typically allow the same function to be launched concurrently on thousands of
threads. This is exactly what we need, since we can update simultaneously
different replicas, and also non-neighboring spins within the same replica,
because the interactions are only between nearest neighbors (see
Sect. \ref{sec-app:parallelization}).

More details on the specific hardware and codes will be given in
Ref.~\onlinecite{Marcotesis}.

\subsection{Parallelization}\label{sec-app:parallelization}

Our update-schemes support two levels of parallelism. Heat-bath and
overrelaxation are parallelized within a single lattice. On the other hand,
parallel tempering concerns $2 N_T$ independent lattices (two replicas, see
Sect.~\ref{sec:obs}, at $N_T$ temperatures). Clearly, spins in different
lattices can be updated simultaneously (between temperature swaps). For small
system sizes, the $2N_T$ lattices can be updated efficiently within a single
GPU. Yet, for $L=64$ we have found it convenient
to speed up by employing $N_T$ GPUs, each of them simulating two lattices.

\subsubsection{Single-GPU}
Our parallelization scheme was not very different from the one described
extensively in previous works such as Refs. \onlinecite{Bernaschi11}
and  \onlinecite{Yavorskii12}, so we limit ourselves to remark
that we used binary couplings in order to be able to store a full coupling in a 
single byte. Also,
due to the fact that the lattice positions were evaluated with bitwise operations,
and to our coalesced memory-reading scheme, \cite{Marcotesis}
our program was mostly efficient when the size of the lattice was a power 
of two, so we favored simulations on those sizes.

\subsubsection{Multi-GPU}
For $L=64$ and $D=0.5$ the relaxation times were too long to be able to thermalize on 
a single GPU.
Therefore, we prepared a code that mixed CUDA
and MPI, in order to be able to concentrate a major computing capability on a
single sample.  We took advantage of the two levels of parallelization that
our update algorithms allow. We
used $N_\mathrm{GPU}=N_T=45$ GPUs, each updating only two independent lattices
with the same couplings, but not necessarily with the same temperature.  At
the level of the single GPU, the way we swept the lattice with heat-bath and
overrelaxation was similar to the single-GPU version. Yet, we had to arrange
it in order to get the same thread occupancy as in the single-GPU version.
Our choice has been to divide the lattice in rows of 8 spins
along the $x$ axis. Non-neighboring rows were updated at the same time. A side
advantage of this scheme was that we could use for it the same type of
coalesced memory reading that we developed for the single-GPU lattice
sweeping.

\begin{figure}[ht]
  \includegraphics[angle=270, width=\columnwidth]{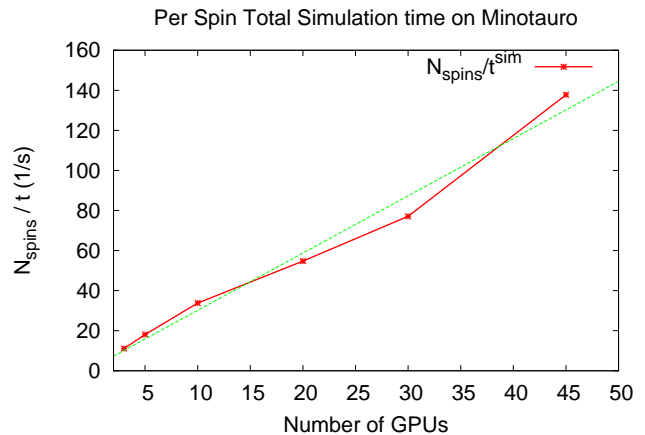}
  \caption{(color online). Scaling of the computing time with the number of GPUs
    $N_\mathrm{GPU}$. Benchmark performed on the \emph{Minotauro} GPU
    cluster. \cite{bsc}}
  \label{fig:scalingGPU}
\end{figure}

This arrangement resulted in an extremely small overhead when passing from the
single to the multiple-GPU algorithm. We were also favored by other factors.
Parallel tempering only requires the exchange with the master of a double
precision number.  Also, the long correlation times allow to take
measurements with low frequency. As a consequence of all this, we obtained a
linear scaling of the computing time with the number of GPUs, $N_T$
(Fig. \ref{fig:scalingGPU}).

\subsection{Pseudo-Random Number Generator}\label{sec-app:PRNG}

Pseudo-random number generators (PRNGs) are a critical issue in the
implementation of stochastic algorithms, \cite{Knuth81} but even more in cases
like ours, where each of the $N_\mathrm{threads}$ threads had to carry its own
PRNG, and we had a large number of them acting in parallel on the same
lattice. This became a major problem especially in the simulations with MPI,
where a huge number of PRNGs was concentrated on only two lattices. It was
crucial to guarantee the statistical independence of the $N_\mathrm{threads}$
pseudo-random sequences. We consider three different aspects: ($a$) the PRNG
that each thread uses, ($b$) the initialization of the generators and ($c$)
our tests on the generators.

\subsubsection{The generator}
We resorted to a linear combination of Parisi-Rapuano with congruential
generators. \cite{Yllanes09}

With the Parisi-Rapuano sequence, \cite{Parisi85} the $n^\mathrm{th}$ pseudo-random number $P_n$ is generated through the following
relations:
\begin{eqnarray}
  y_n &=& (y_{n-24} + y_{n-55})\, \mathrm{mod\,2^{64}}\\\nonumber
  P_n &=& y_n ~ \text{XOR} ~ y_{n-61}\,,
\end{eqnarray}
where XOR is the exclusive OR logic operator, and $y_i$ are 64-bit unsigned
integers.  Although some pathologies have been found in the 32-bit
Parisi-Rapuano PRNG,\cite{Ballesteros98} it looks like its 64-bit version is
solid. \cite{Fernandez05}

On the other side, we used a 64-bit congruential generator, where the
$n^\mathrm{th}$ element of the sequence, $C_n$, was given
by:\cite{Knuth81,LEcuyer99}
\begin{equation}\label{eq:congruential}
  C_n = (C_{n-1} \times 3202034522624059733 + 1 )\,\mathrm{mod\,2^{64}}\,.
\end{equation}
Also this generator is not reliable when used alone.\cite{Yllanes09,Ossola04}

The final pseudo-random number $R_n$ was obtained by summing $P_n$ and $C_n$:
\begin{equation}\label{eq:ParisiRapuanoCongruencial}
  R_n = (P_n+C_n)\,\mathrm{mod\,2^{64}}\,.
\end{equation}

\subsubsection{Initializing the generators}

We have found that problems arise if special care is not devoted to the
initialization of the random numbers. This is particularly important in the
case of multiple GPUs where $N_\mathrm{threads}=32768$ threads concurrently
update the spins in only two lattices.

We decided to use one seed per node. This seed was used to initialize a 64-bit
Congruential PRNG, Eq.~\eqref{eq:congruential}. We employed it to initialize
the state vector of 24-bit Luescher PRNG.\cite{Luescher94} The 24-bits words
were obtained from three consecutive congruential calls (we kept the most
significant byte from each call). As for the Luescher generator, we employed
the \emph{full luxury} version, which is fireproof but slow. We took the 8
most significant bits from each Luescher call to fill up the state vector of
the 64-bit PRNGs in Eq.~\eqref{eq:ParisiRapuanoCongruencial}. We were probably
excessively cautious, given the high quality of the full-luxury generator, but
initialization takes only a small fraction of the total computing time.

\subsubsection{Tests}
We tested with success our random sequences through the whole battery of tests proposed by Marsaglia in Ref.~\onlinecite{Marsaglia}.
To be sure the sequences were reliable also with concurrent threads, we also generated $N_\mathrm{threads}$
sequences and tested them \emph{horizontally}, i.e. taking first the first number of each sequence,
then the second, and so on.

Also, we made simulations with ferromagnetic couplings demanding the energies to be equal,
up to the $7^\mathrm{th}$ significant digit, to those obtained with an independent
CPU program.

Finally, it has been pointed out that local Schwinger-Dyson relations (see
e.g. Ref. \onlinecite{Rivers90}) can be useful to assess the quality of
PRNGs.\cite{Ballesteros98} The relevant identity here is
\begin{equation}
  2 T \left\langle \vec s_\xb\cdot \vec h_\xb \right\rangle - \left\langle (\vec h_\xb)^2 - (\vec s_\xb\cdot \vec h_\xb)^2\right\rangle = 0 \,.
\end{equation}
We averaged it over all the sites in the lattice, in order to obtain a more
stringent test for the simulations.


\begin{thebibliography}{unsrt}
\bibitem{scalingRelation} Recall that $\gamma_\mathrm{CG} = \nu (2 -\eta_\mathrm{CG})$ where $\gamma_\mathrm{CG}$ is the critical index for the CG susceptibility, while $\nu$ is the correlation-length exponent.
\bibitem{Independence} Independence from microscopic details such as the
  disorder distribution has been found for spin glasses,\cite{Hasenbusch08,
    Jorg06, Katzgraber06, Jorg08} as well as for other disordered systems like
  the Random Field Ising model,\cite{Fytas13} or disordered ferromagnets
  (either site~\cite{Ballesteros98b} or bond~\cite{Berche04,Malakis12}
  diluted).
\bibitem{generalize} It is enough to define the local field as $\vec{h_\xb} = \sum_{\Vert{\vn x}-{\vn y}\Vert=1}[J_{\xb\yb} \vec s_\yb + D_{\xb\yb} \vec s_{\yb}]$.
\bibitem{covariance} Some of the points we used for those extrapolations
  shared some of the data. For example, the crossing of $\xi$ for $L=8,16$,
  had in common the points from size $L=16$ with the pair $L=16,32$. Hence, in
  the fits we have taken in account the covariance matrix that gave a measure
  of the anticorrelation between measures that share data.
\bibitem{Dequivalence} {In Ref.~\onlinecite{MartinPerez11} the
  $D_{ij}^{\alpha\beta}$ are uniformly distributed between $\pm D=0.05$, while
  in our work we use binary couplings. If we want to compare them, we have to
  use \mbox{$D=1/\sqrt{1200}\simeq0.03$}}.
\bibitem{PDdisclaimer} In the phase diagram we show, the $D=0$ point comes
  from Ref. \onlinecite{Fernandez09}, where chiral and spin glass transition
  are assumed to be coupled.  There is disagreement on whether
  $T_\mathrm{SG}=T_\mathrm{CG}$ also in the isotropic case. Yet, we do plot it
  as a single transition because although $T_\mathrm{SG}$ might be lower than
  $T_\mathrm{CG}$, their best estimates are compatible (and not
  distinguishable in the plot).
\end{thebibliography}
\end{document}